\documentclass[aps,prd,notitlepage,showpacs,nofootinbib,superscriptaddress]{revtex4-1}

\usepackage{graphicx}
\usepackage[utf8]{inputenc} 

\usepackage{amsmath}
\usepackage{amssymb}
\usepackage{float}
\usepackage{comment}
\usepackage{slashed}
\usepackage[normalem]{ulem}
\usepackage{booktabs}
\usepackage{array}

\usepackage{caption}
\usepackage{subcaption}
\usepackage{soul}

\usepackage{amsmath}
\usepackage{amssymb}

\usepackage[usenames,dvipsnames]{color}
\usepackage[colorinlistoftodos]{todonotes}
\usepackage[colorlinks=true,citecolor=darkred,urlcolor=darkred, pdfborder={0 0 0}]{hyperref}
\usepackage[normalem]{ulem}

\definecolor{darkred}{rgb}{0.6,0,0}
\usepackage[colorinlistoftodos]{todonotes}

\definecolor{linkcolor}{rgb}{0,0,0.5}

\newcommand {\ignore}[1]{}

\def\TrTrOne{$\mathrm{SU(3)_c \otimes SU(3)_L \otimes U(1)_X}$ }
\def\TrTrOO{$\mathrm{SU(3)_c \otimes SU(3)_L \otimes U(1)_X \otimes U(1)_N}$ }

 \def\three{\ensuremath{\mathbf{3}}}
 \def\threeS{\ensuremath{\mathbf{3^*}}}

\def\gsim{\raise0.3ex\hbox{$\;>$\kern-0.75em\raise-1.1ex\hbox{$\sim\;$}}}
\def\lsim{\raise0.3ex\hbox{$\;<$\kern-0.75em\raise-1.1ex\hbox{$\sim\;$}}}

\newcommand{\sm}{{Standard Model }}

\usepackage{soul}

\definecolor{mightnightblue}{RGB}{25,25,112}

\definecolor{brown}{rgb}{0.59, 0.29, 0.0}

\def\21{$\mathrm{SU(2)_L \otimes U(1)_Y}$}

\def\sm{standard model }

\newcommand{\AddrAHEP}{%
  AHEP Group, Institut de F\'{i}sica Corpuscular --
  C.S.I.C./Universitat de Val\`{e}ncia, Parc Cient\'ific de Paterna.\\
 C/ Catedr\'atico Jos\'e Beltr\'an, 2 E-46980 Paterna (Valencia) - SPAIN}

\begin{document}


\title{\boldmath \color{BrickRed} Dark matter stability from Dirac neutrinos in scotogenic 3-3-1-1 theory }

\author{Julio Leite}\email{julio.leite@ific.uv.es}
\affiliation{\AddrAHEP}
\affiliation{Centro de Ci\^encias Naturais e Humanas, Universidade Federal do ABC,\\ 09210-580, Santo Andr\'e-SP, Brasil}

\author{América Morales}\email{america@fisica.ugto.mx }
\affiliation{Departamento de F\'isica, DCI, Campus Le\'on, Universidad de
Guanajuato, Loma del Bosque 103, Lomas del Campestre C.P. 37150, Le\'on, Guanajuato, M\'exico}

\author{Jos\'{e} W. F. Valle}\email{valle@ific.uv.es}
\affiliation{\AddrAHEP}

\author{Carlos A. Vaquera-Araujo}\email{vaquera@fisica.ugto.mx}
\affiliation{Consejo Nacional de Ciencia y Tecnolog\'ia, Av. Insurgentes Sur 1582. Colonia Cr\'edito Constructor, Del. Benito Ju\'arez, C.P. 03940, Ciudad de M\'exico, M\'exico}
\affiliation{Departamento de F\'isica, DCI, Campus Le\'on, Universidad de
Guanajuato, Loma del Bosque 103, Lomas del Campestre C.P. 37150, Le\'on, Guanajuato, M\'exico}

\begin{abstract}
\vspace{0.5cm}

We propose the simplest TeV-scale scotogenic extension of the original 3-3-1 theory, where dark matter stability is linked to the Dirac nature of neutrinos, which results from an unbroken $B-L$ gauge symmetry. The new gauge bosons get masses through the interplay of spontaneous symmetry breaking \textit{\`a la Higgs} and the Stueckelberg mechanism.

\end{abstract}

\maketitle
\noindent


\section{Introduction}
\label{Sect:intro}

Despite its amazing phenomenological success, almost no one thinks of the \sm as the final theory, so many are its drawbacks.
Amongst these, the issues of neutrino mass, dark matter, the number of families and the strong CP problem stand out as important items in the wish list of theorists.
Here, we propose a \sm extension where these appear closely interconnected.
To do this, we build up upon a minimal gauge extension of the original Singer-Valle-Schechter (SVS) 3-3-1 model \cite{Singer:1980sw}.
This was the first electroweak extension of the \sm in which the existence of three families of quarks and leptons is closely related to anomaly cancellation.
Indeed, in this \TrTrOne theory, one assumes that leptons transform as $\mathrm{SU(3)_L}$ anti-triplets, while two families of left-handed quarks
transform as triplets and the last one is an anti-triplet.
This choice comes from anomaly cancellation and once adopted, leads to the prediction of three families of quarks and leptons~\cite{Singer:1980sw,Pisano:1991ee,Frampton:1992wt,Foot:1994ym}.
In order to make the construction as minimal as possible, we also adopt the choice made in~\cite{Valle:1983dk} of identifying the third component of the leptons as a ``right-handed'' neutrino, so that neutrinos are of a Dirac nature and their masses are generated at tree-level.
However this early formulation is not compatible with the current neutrino oscillation data~\cite{deSalas:2017kay}, as it predicts one massless and two mass-degenerate neutrinos. 
Besides, an unaesthetic feature of this construction is that lepton number symmetry emerges in SVS as a combination of a gauge symmetry and a global one.

In what follows, we explore a simple scheme with a viable neutrino spectrum and realize the scotogenic dark matter paradigm~\cite{Ma:2006km}, which postulates that neutrino masses arise through the radiative exchange of a ``dark matter'' sector.
The idea of relating dark matter stability to the Diracness of neutrinos has been proposed in~\cite{Chulia:2016ngi}, employing residual discrete symmetries arising from the partial breaking of a global $B-L$ symmetry~\cite{Chulia:2016ngi,CentellesChulia:2017koy,Hirsch:2017col,CentellesChulia:2018bkz,CentellesChulia:2018gwr,Bonilla:2018ynb,Ma:2019coj}.
  An alternative proposal to link Dirac neutrino masses and dark matter stability is through a fully conserved global $B-L$ symmetry.
This idea has been pursued in the context of bound-state dark matter~\cite{Reig:2018mdk}, in which the radiative generation of Dirac neutrino masses is mediated by the exchange of bound-state-dark-matter \textit{constituents}.

In this paper, we choose a different route, namely, a scenario where dark matter stability is interconnected to the Diracness of neutrinos in the framework of a dynamical theory with gauged lepton number.
In order to achieve this, we build upon a minimally extended class of SVS theories developed within the framework of the \TrTrOO gauge symmetry~\cite{Dong:2013wca,Alves:2016fqe,Dong:2018aak,Kang:2019sab,Leite:2019grf}.
The extra Abelian $\mathrm{U(1)_N}$ group allows a consistent embedding of $B-L$ as a fully dynamical gauge symmetry.
We propose a simple 3-3-1-1 extension of the original SVS theory, where neutrino masses arise ``scotogenically'' (i.e. through the exchange of ``dark'' particles) at the one-loop level.
The unbroken $B-L$ symmetry acts as the protecting symmetry responsible for both neutrino Dirac masses and stabilization of dark matter.
Such a conserved gauge symmetry is shown to be fully consistent, as we adopt the Stueckelberg mechanism to provide a mass to the associated gauge field while keeping the symmetry intact.
Our present construction follows the path of the simple Stueckelberg \cite{Ruegg:2003ps} $\mathrm{U(1)}$ extension of the standard model proposed in Ref.~\cite{Leite:2020wjl}.
However, this is achieved within a richer framework that provides not only a dynamical realization of the proposal that dark matter stability and Diracness are closely inter-related,
but also touches other \sm shortcomings such as the number of families and the strong CP problem. In particular, the existence of three families of quarks and leptons is linked to anomaly cancellation.
Our present model also provides an example of ``predestined'' dark matter~\cite{Ma:2018bjw}, in the sense that the specific quantum numbers of the new fermion and scalar multiplets
automatically ensure the existence of a stable dark matter candidate, without the \textit{ad hoc} imposition of any additional symmetry.

The paper is organized as follows.
 In Sec.~\ref{sec:model}, we define the proposed model in terms of its field content and symmetries.
 The scalar sector is studied in Sec.~\ref{sec:scalar}. 
 In Sec.~\ref{sec:gauge}, we derive the extended electroweak vector boson spectrum taking into account contributions coming from both the spontaneous symmetry and Stueckelberg mechanisms. 
 The charged fermion spectrum is presented in Sec.~\ref{sec:fermion}, the scotogenic neutrino masses are calculated in Sec.~\ref{sec:neutrino} and in Sec.~\ref{sec:DM} we study the case for a complex dark matter scalar candidate.
 Finally, the conclusions are presented in Sec. \ref{sec:Conclusions}.

\section{The Model}
\label{sec:model}

In the present model, not only the Abelian electromagnetic symmetry $\mathrm{U(1)_Q}$ but also the $\mathrm{U(1)_{B-L}}$ symmetry emerges as a conserved residual subgroup of the \TrTrOO gauge symmetry, or 3-3-1-1 for short.
In 3-3-1-1 models, the electric charge operator can be generically written as
\begin{equation}
Q=T^3 +\beta T^8 +X,
\end{equation}
while the $B-L$ generator is expressed as
\begin{equation}
B-L=\beta^\prime T^8 +N,
\end{equation}
with $T^{a}$ $(a=1,\dots,8)$, $X$ and $N$ as the respective generators of $\mathrm{SU(3)_L}$, $\mathrm{U(1)_X}$ and $\mathrm{U(1)_N}$ \cite{Dong:2015yra}.
The choices of the constants $\beta$ and $\beta^\prime$ define different versions of the model, and for the SVS model, we have $\beta=1/\sqrt{3}$ and $\beta^\prime=4/\sqrt{3}$.
This specific choice ensures the $B-L$ assignment in the SVS model with its original field content is anomaly free and can be promptly promoted to local.
On the other hand, other $\beta^\prime$ values would require new fermions to cancel the $B-L$ anomalies\footnote{Explicit calculation of anomaly coefficients for generic $\beta$ and $\beta^\prime$ can be found in Ref. \cite{Dong:2015yra}.}; see, {\it e.g.} Refs. \cite{Dong:2013wca}.
Here, we stick to the SVS choice given in Table \ref{tab1}. This gives all the quantum number assignments for the fields contained in our model.
In addition to the fields present in the original SVS model, we have included three two-component Majorana fermion singlets $S_R^a$, $a=1,2,3$ and one scalar anti-triplet $\Phi_4$.
Notice that the Majorana fermions are full gauge singlets and hence, carry no anomaly.
The global $\mathrm{U(1)_{PQ}}$ symmetry forbids the term $\overline{(\psi^{a}_L)^c} \,\Phi_1\, \psi^{b}_L $, which appears in Ref.~\cite{Valle:1983dk} and leads to tree-level Dirac neutrino masses.
However, as it will be discussed in Sec. \ref{sec:scalar}, this global symmetry is softly broken in the scalar sector, by the trilinear $\Phi_1 \Phi_2 \Phi_3$ coupling.
As we will see, this avoids the disastrous presence of a visible axion field.
\begin{table}[h]
	\begin{centering}
     {\renewcommand{\arraystretch}{1.7}
		\begin{tabular}{ccccc}
            \hline
			\toprule 
			\,\,Field\,\, & 3-3-1-1 rep  & Components & $B-L$ & \,\,$\mathrm{U(1)_{PQ}}$\,\,
			\tabularnewline
			\hline\hline
			\midrule
			$Q_{\alpha L}$ & $\left(\mathbf{3},\three,0,-\frac{1}{3}\right)$ & $\left( (u_{\alpha L},d_{\alpha L}),D_{\alpha L}\right)^{T}$ & $\left(\frac{1}{3},\frac{1}{3},-\frac{5}{3}\right)^{T}$ & $1$\tabularnewline
			$Q_{3L}$ & $\left(\mathbf{3},\threeS,\frac{1}{3},1\right)$ & $\left( (b_L,-t_L),U_{3L}\right)^{T}$ & $\left(\frac{1}{3},\frac{1}{3},\frac{7}{3}\right)^{T}$& $1$ \tabularnewline
			$u_{a R}$ & $\left(\mathbf{3},\mathbf{1},\frac{2}{3},\frac{1}{3}\right)$ & $u_{a R}$ & $\frac{1}{3}$ & $4$\tabularnewline
			$ U_{3R}$ & $\left(\mathbf{3},\mathbf{1},\frac{2}{3},\frac{7}{3}\right)$ & $ U_{3R}$ & $\frac{7}{3}$ & $4$\tabularnewline
			$d_{a R}$ & $\left(\mathbf{3},\mathbf{1},-\frac{1}{3},\frac{1}{3}\right)$& $d_{a R}$ & $\frac{1}{3}$ & $-2$\tabularnewline
            $D_{\alpha R}$ & $\left(\mathbf{3},\mathbf{1},-\frac{1}{3},-\frac{5}{3}\right)$ & $D_{\alpha R}$ & $-\frac{5}{3}$ & $-2$\tabularnewline
            $\psi_{aL}$ & $\left(\mathbf{1},\threeS,-\frac{1}{3},-\frac{1}{3}\right)$ & $\left( (e_{aL},-\nu_{aL}),\nu_{aR}^{c}\right)^{T}$ & $\left(-1,-1,+1\right)^{T}$& $-3$ \tabularnewline
			$e_{aR}$ & $\left(\mathbf{1},\mathbf{1},-1,-1\right)$ & $e_{aR}$ & $-1$ & $-6$\tabularnewline
            \hline
            $S_{aR}$ & $\left(\mathbf{1},\mathbf{1},0,0\right)$ & $S_{aR}$ & $0$ & $0$\tabularnewline
            \hline \hline
			$\Phi_{1}$ & $\left(\mathbf{1},\threeS,\frac{2}{3},\frac{2}{3}\right)$ & $\left(\left(\phi_{1}^{0},-\phi_{1}^{+}\right),\widetilde{\phi}_{1}^{+}\right)^{T}$ & $\left(0,0,2\right)^{T}$ & $3$\tabularnewline
			$\Phi_2$ & $\left(\mathbf{1},\threeS,-\frac{1}{3},\frac{2}{3}\right)$ & $\left(\left(\phi_{2}^{-},-\phi_{2}^{0}\right),\widetilde{\phi}_{2}^{0}\right)^{T}$ & $\left(0,0,2\right)^{T}$  & $-3$\tabularnewline
			$\Phi_3$ & $\left(\mathbf{1},\threeS,-\frac{1}{3},-\frac{4}{3}\right)$ & $\left(\left(\phi_{3}^{-},-\phi_{3}^{0}\right),\widetilde{\phi}_{3}^{0}\right)^{T}$ & $\left(-2,-2,0\right)^{T}$ & $-3$ \tabularnewline
			\hline
			\,\,\,$\Phi_4$\,\,\, & \,\,\,$\left(\mathbf{1},\threeS,-\frac{1}{3},-\frac{1}{3}\right)$\,\,\, & \,\,\,$\left(\left(\phi_{4}^{-},-\phi_{4}^{0}\right),\widetilde{\phi}_{4}^{0}\right)^{T}$ \,\,\,& \,\,\,$\left(-1,-1,1\right)^{T}$\,\,\, & $-3$
\tabularnewline
			\bottomrule
			\hline
		\end{tabular}}
		\par\end{centering}
		\caption{Field content and symmetry transformations.}\label{tab1}
\end{table}
Notice also that, since $B-L$ remains unbroken, the matter parity subgroup, generated by the matter parity $M_P = (-1)^{3(B-L)+2s}$, where $s$ is the field's spin, is also fully preserved.
Under $M_P$, all the fields in the original SVS model transform trivially, whereas the new fields transform as $(S_{aR}, \Phi_4)\xrightarrow{M_P}-(S_{aR}, \Phi_4)$.
Therefore, the lightest among the $M_P$-odd fields is stable, and, if electrically neutral, it can play the role of dark matter.

\section{Scalar Sector}
\label{sec:scalar}

Our model contains four triplet scalars, three of them are Higgs-like, even under matter parity, while $\Phi_4$ is ``dark'' or $M_P$ odd.
The resulting scalar potential is given by  
\begin{eqnarray}\label{V1}
V_{\rm \Phi} &=& \,
 \sum_{i=1}^{4}\left[\mu_i^2 \Phi^{\dagger}_i\Phi_i+\lambda_{i}(\Phi^{\dagger}_i\Phi_i)^2\right]+\sum_{i<j}\left[\lambda_{ij}(\Phi^{\dagger}_i\Phi_i)(\Phi^{\dagger}_j\Phi_j)+\tilde{\lambda}_{ij}(\Phi^{\dagger}_i\Phi_j)(\Phi^{\dagger}_j\Phi_i)
\right]\,\nonumber\\
 & & +\left(-\frac{\mu_{\phi}}{\sqrt{2}}\Phi_1\Phi_2\Phi_3+\frac{\lambda^{\prime}}{2}\Phi_2^{\dagger}\Phi_4\Phi_3^{\dagger}\Phi_4+\mathrm{h.c.}\right),
\end{eqnarray}
where the cubic term characterized by $\mu_{\phi}$ breaks the $\mathrm{U(1)_{PQ}}$ symmetry softly. 

The scalar multiplets are decomposed as
\begin{equation}
\label{scalar3plets3}
\Phi_1=\left(
\begin{array}{c}
\frac{v_1+s_1+i a_1}{\sqrt{2}}\\
-\phi^{+}_{1}\\
\widetilde{\phi}^{+}_{1}
\end{array}
\right),\qquad
\Phi_2=\left(
\begin{array}{c}
\phi_2^{-}\\
\frac{v_2-s_2-i a_2}{\sqrt{2}}\\
\widetilde{\phi}_2^0
\end{array}
\right),\qquad
\Phi_3=\left(
\begin{array}{c}
 \phi_3^{-} \\
 -\phi_3^0 \\
 \frac{w+s_3+i a_3}{\sqrt{2}} \\
\end{array}
\right),\qquad
\Phi_4=\left(
\begin{array}{c}
 \phi_3^{-} \\
 -\phi_4^0 \\
\widetilde{\phi}_4^{0} \\
\end{array}
\right),
\end{equation}
where $v_1/\sqrt{2}, v_2/\sqrt{2}$ and $w/\sqrt{2}$ represent vacuum expectation values (vevs), with $w^2\gg v_1^2+v_2^2\equiv v^2_{EW}$.
Notice that, with the assumed vev alignment, the $B-L$ symmetry remains conserved, and the minimization of the potential leads to the tadpole equations,
\begin{equation}
\begin{split}
 v_1\left(2 \mu^2_1+2 \lambda _1 v_1^2+\lambda _{12}
   v_2^2+\lambda _{13}  w^2\right)- v_2 w \mu _{\phi }=0,\\
v_2\left(2 \mu^2_2+2 \lambda _2 v_2^2+\lambda _{12}
    v_1^2+\lambda _{23}  w^2\right)- v_1 w \mu _{\phi }=0,\\
w\left(2 \mu^2_3+2 \lambda_3 w^2+\lambda_{13}v_1^2+\lambda _{23}v_2^2\right)- v_1 v_2 \mu _{\phi
   }=0,      
\end{split}
\end{equation}
which can be simultaneously solved for $\mu^2_1$, $\mu^2_2$ and $\mu^2_3$. 
In the following subsections, we present the physical states of the scalar sector and their respective masses. 

\subsection{CP-even scalars}

After spontaneous symmetry breaking, the CP-even components of the fields that acquire a vev mix according to the following squared mass matrix, written in the basis $(s_1,\, s_2, \,s_3)$:
\begin{equation}
\begin{split}
M^2_{s_1,\, s_2, \,s_3}&=
\left(
\begin{array}{ccc}
 2 v_1^2 \lambda _1+\frac{v_2 w \mu _{\phi }}{2v_1} & - v_1 v_2
   \lambda _{12}+\frac{w \mu _{\phi }}{2} &  v_1 w \lambda _{13}-\frac{v_2 \mu
   _{\phi }}{2} \\
  - v_1 v_2
   \lambda _{12}+\frac{w \mu _{\phi }}{2} & 2  v_2^2 \lambda
   _2+\frac{v_1 w \mu _{\phi }}{2v_2} & - v_2 w \lambda _{23}+\frac{v_1 \mu
   _{\phi }}{2} \\
  v_1 w \lambda _{13}-\frac{v_2 \mu
   _{\phi }}{2} & - v_2 w \lambda _{23}+\frac{v_1 \mu
   _{\phi }}{2} & 2 w^2 \lambda _3 +\frac{v_1 v_2 \mu
   _{\phi }}{2w} \\
\end{array}
\right)\,.
\end{split}
\end{equation}
Diagonalization yields three physical mass-eigenstate scalars. Assuming for simplicity, the hierarchy $\mu_\phi,\,w\gg v_1,\, v_2$, the lightest one can be identified with the \sm Higgs boson discovered at the LHC, 
\begin{equation}
h\approx\frac{v_2 s_2-v_1 s_1}{\sqrt{v_1^2+v_2^2}}. 
\end{equation}
Its squared mass is given as
\begin{equation}
m_h^2\approx\left(2 \lambda _1-\frac{\lambda _{13}^2}{2 \lambda _3}\right)
   v_1^2+v_2^2 \left(2 \lambda _{12}-\frac{\lambda _{13} \lambda _{23} }{\lambda _3 }-\frac{\mu _{\phi
   }^2}{\lambda _3 w^2}\right)+\left(2 \lambda _2-\frac{\lambda _{23}^2}{2 \lambda
   _3}\right)\frac{v_2^4}{v_1^2}+\frac{ \mu _{\phi } \left(\lambda _{13} v_1^2 v_2+\lambda _{23}
   v_2^3\right)}{\lambda _3 v_1 w},
\end{equation}
where all parameters, other than  $\mu_\phi$ and $w$, lie at the electroweak scale.
The remaining scalars are heavy and can be approximately identified as
\begin{equation}\label{Hmass}
\begin{split}
H_1&\approx\frac{v_2 s_1+v_1 s_2}{\sqrt{v_1^2+v_2^2}},\qquad m_{H_1}^2\approx\frac{\left(v_1^2+v_2^2\right) w \mu _{\phi }}{2v_1 v_2},\\
H_2&\approx s_3,                                     \qquad\qquad\qquad m_{H_2}^2\approx 2 w^2 \lambda _3.\\
\end{split}
\end{equation}
In principle,  $\mu_\phi$ can be even lower than the electroweak scale. In that case, this sector would give rise to two light scalars and a heavy one. In what follows, we assume an arbitrary $\mu_\phi$ scale and a VEV hierarchy $w\gg v_1,\, v_2$.

\subsection{CP-odd scalars}

Similar to the CP-even scalars, the CP-odd components also mix through the squared mass matrix, 
\begin{equation}
\begin{split}
M^2_{a_1, \, a_2,  \, a_3}&=\frac{\mu_{\phi}}{2}\left(
\begin{array}{ccc}
 \frac{v_2 w }{v_1} & -w  &  v_2   \\
  -w  &  \frac{v_1 w }{v_2} &  -v_1 \\
  v_2   &   -v_1   & \frac{v_1 v_2}{w} \\
\end{array}
\right)\,,
\end{split}
\end{equation}
in the basis $(a_1, \, a_2,  \, a_3)$. Upon diagonalization, we find two Nambu-Goldstone bosons that can be identified as
\begin{equation}
G_1=\frac{v_1 a_1+v_2 a_2}{\sqrt{v_1^2+v_2^2}}\,,\,\,\,\,\,\,\,\,\,\,\,\,G_2=\frac{-v_1 a_1+wa_3}{\sqrt{v_1^2+ w^2}},
\end{equation}
and one physical pseudoscalar,
\begin{equation}\label{pseudo}
A'_1=\frac{v_2 w a_1-v_1 w a_2+v_1 v_2 a_3}{\sqrt{v_1^2v_2^2+v_1^2w^2+v_2^2
   w^2}},
\end{equation}
with a squared mass
\begin{equation}\label{pseudo2}
m^2_{A'_1}=\frac{\mu _{\phi } \left(v_1^2v_2^2+v_1^2w^2+v_2^2
   w^2\right)}{2 v_1 v_2 w}.
\end{equation}
The importance of the $\mathrm{U(1)_{PQ}}$ soft-breaking term characterized by $\mu_\phi$ can be better understood by looking at the equations above and Table \ref{tab1}. 
In the limit $\mu_{\phi}\to 0$, $\mathrm{U(1)_{PQ}}$ becomes a classical global symmetry of the model, whose spontaneous breaking by the vevs of the scalar fields
would imply the existence of a massless Nambu-Goldstone boson, namely, the pseudoscalar defined in Eq. (\ref{pseudo}).
However, the Peccei-Quinn-like symmetry has an associated $[SU(3)_C]^2 \mathrm{U(1)_{PQ}}$ anomaly.
Therefore, instead of a massless field, we would have a pseudo-Goldstone boson, an axion field getting its mass via nonperturbative effects.
The existence of such a ``low-scale'' axion ($w=10$ TeV), {\it \`a la} Weinberg-Wilczek \cite{Weinberg:1977ma,Wilczek:1977pj}, is ruled out phenomenologically,
as noted in Ref. \cite{Pal:1994ba}.
  Alternative 3-3-1 proposals including gauge singlet scalars with non-vanishing $\mathrm{U(1)_{PQ}}$ charges have been considered~\cite{Dias:2003iq, Dias:2018ddy}.
  This way one can make the axion invisible and thus, viable by introducing a large $\mathrm{U(1)_{PQ}}$ breaking scale.
 Here we do not follow this path. Instead, we avoid the presence of the visible axion simply by softly breaking $\mathrm{U(1)_{PQ}}$ via the trilinear $\Phi_1\Phi_2\Phi_3$ term, instead of adding more scalars. Apart from minimality, this also ensures that tree-level neutrino masses are absent. 

\subsection{Complex neutral scalars}

The complex neutral scalars that do not acquire vevs can be grouped in pairs according to their $B-L$ charges, as follows. 
First notice that, since $B-L$ is conserved, only fields with the same $B-L$ charges can mix.
Since the fields $\widetilde{\phi}_2^0$ and $\phi_3^{0}$ carry opposite $B-L$ charges, we define a $B-L=2$ basis as $(\widetilde{\phi}_2^0,\, \phi_3^{0*})$.
In this basis, we can write down the following squared mass matrix 
\begin{equation}
\begin{split}
M^2_{\widetilde{\phi}_2^0, \phi_3^{0}}&=\frac{1}{2}\left(
\begin{array}{cc}
 w \left(w \tilde{\lambda} _{23}+\frac{v_1 \mu _{\phi }}{v_2}\right) &
   - v_1 \mu _{\phi }-v_2 w \tilde{\lambda} _{23} \\- v_1 \mu _{\phi }-v_2 w \tilde{\lambda}_{23} & v_2 \left(v_2 \tilde{\lambda}
   _{23}+\frac{ v_1 \mu _{\phi }}{w}\right) \\
\end{array}
\right)\,.
\end{split}
\end{equation}
Upon diagonalization, we find a massless complex scalar, shown in the next section to be absorbed \textit{\`a la Goldstone} by the gauge sector, 
\begin{equation}\label{G3}
G_3=\frac{v_2 \widetilde{\phi}_2^0+w \phi_3^{0*}}{\sqrt{v_2^2+w^2}},
\end{equation}
and a heavy complex scalar field,
\begin{equation}
\varphi=\frac{-w \widetilde{\phi}_2^0+v_2 \phi_3^{0*}}{\sqrt{v_2^2+w^2}},\qquad m^2_{\varphi}=\frac{(v_2^2+w^2)(\tilde{\lambda}_{23}v_2w+v_1\mu_{\phi})}{2v_2w}.
\end{equation}

Likewise, coming to the remaining fields, these can be grouped in a basis with $B-L=1$ as $(\widetilde{\phi}_4^0,\,\phi_4^{0*})$. The corresponding squared mass matrix is
\begin{equation}\label{MsqS}
\begin{split}
M^2_{\widetilde{\phi}_4^0,\,\phi_4^0}&=\frac{1}{2}\left(
\begin{array}{cc}
 v_1^2 \lambda_{14}+v_2^2 \lambda_{24}+w^2(\lambda_{34}+\tilde{\lambda}_{34})+2\mu^2_4  &- \frac{1}{2} \lambda^{\prime} v_2w    \\
- \frac{1}{2} \lambda^{\prime} v_2w  &  v_1^2 \lambda_{14}+v_2^2(\lambda_{24}+\tilde{\lambda}_{24})+w^2 \lambda_{34}+2\mu^2_4\\
\end{array}
\right)\,,
\end{split}
\end{equation}
that can be diagonalized as
\begin{equation}
\left(
\begin{array}{c}
\eta_1\\
\eta_2  \\
\end{array}
\right)=\left(
\begin{array}{cc}
\cos{\theta} & \sin{\theta}  \\
-\sin{\theta}  &  \cos{\theta}   \\
\end{array}
\right) \left(
\begin{array}{c}\widetilde{\phi}^0_4 \\
\phi_4^{0*}
\end{array}
\right)\,,
\end{equation}
with 
\begin{equation}\label{tmix}
\tan{2\theta}=\frac{v_2 w \lambda^{\prime }}{v_2^2 \widetilde{\lambda} _{24}-w^2 \widetilde{\lambda}_{34}}. 
\end{equation}
The physical neutral fields $\eta_1$ and $\eta_2$ defined above have squared masses
\begin{equation}\label{meta}
\begin{split}
m_{\eta_1,\eta_2}^2 =&\frac{1}{4} \left[4 \mu^2_4+2 \lambda _{14}
   v_1^2+v_2^2 \left(2 \lambda _{24}+\tilde{\lambda}_{24}\right)+w^2 \left(2 \lambda
   _{34}+\tilde{\lambda}_{34}\right)\mp\mathcal{F}\sqrt{ \lambda
   ^{\prime 2}v_2^2 w^2+\left(\tilde{\lambda}_{24} v_2^2-\tilde{\lambda}_{34} w^2\right){}^2}\right],\end{split}
\end{equation}
where $\mathcal{F}=\mathrm{sign}(\tilde{\lambda}_{24} v_2^2-\tilde{\lambda}_{34} w^2)$.

\subsection{Charged scalars}

Again, for the charged scalars too, mixing takes place amongst those with the same $B-L$ charges, and they can be separated into three groups. 

The basis $(\phi_1^{\pm},\phi_2^{\pm})$ puts together the charged fields with $B-L=0$, which mix according to the squared mass matrix
\begin{equation}
\begin{split}
M^2_{\phi_1^{\pm},\phi_2^{\pm}}&=\frac{1}{2}\left(
\begin{array}{cc}
v_2\left(\frac{w\mu_{\phi}}{v_1}+ \tilde{\lambda}_{12}v_2\right) & -\mu_{\phi}w- \tilde{\lambda}_{12} v_1 v_2  \\
  -\mu_{\phi}w-\tilde{\lambda}_{12}v_1 v_2 & v_1\left(\frac{w\mu_{\phi}}{v_2}+ \tilde{\lambda}_{12}v_1\right)  \\
\end{array}
\right)\,.
\end{split}
\end{equation}
Upon diagonalization, we find a (complex) ``Goldstone'' boson
\begin{equation}\label{G4}
G^{\pm}_4=\frac{v_1 \phi_1^{\pm}+v_2 \phi_2^{\pm}}{\sqrt{v_1^2+v_2^2}},
\end{equation}
and a massive electrically charged physical scalar,
\begin{equation}
H^{\pm}_1= \frac{-v_2 \phi^{\pm}_1+v_1 \phi^{\pm}_2}{\sqrt{v_1^2+v_2^2}},\qquad m^2_{H^{\pm}_1}=\frac{(v_1^2+v_2^2)(w\mu_{\phi}+v_1v_2\tilde{\lambda}_{12})}{2v_1v_2}.
\end{equation}

The charged scalars with $B-L=\pm2$ are characterized by the following squared mass matrix, written in the basis $(\widetilde{\phi}_1^\pm,\phi^\pm_3)$:
\begin{equation}
\begin{split}
M^2_{\widetilde{\phi}_1^{\pm},\phi_3^{\pm}}&=\frac{1}{2}\left(
\begin{array}{cc}
 w \left(w \tilde{\lambda} _{13}+\frac{ v_2 \mu _{\phi }}{v_1}\right) & v_1 w
   \tilde{\lambda} _{13}+ v_2 \mu _{\phi } \\
 v_1 w \tilde{\lambda} _{13}+ v_2 \mu _{\phi } & v_1 \left(v_1 \tilde{\lambda}
   _{13}+\frac{ v_2 \mu _{\phi }}{w}\right) \\
\end{array}
\right)\,,
\end{split}
\end{equation}
from which one can identify another pair of charged Goldstones,
\begin{equation}\label{G5}
G^{\pm}_5=\frac{-v_1 \widetilde{\phi}_1^{\pm} + w\phi^{\pm}_3}{\sqrt{v_1^2 + w^2}},
\end{equation}
and the heavy charged states,
\begin{equation}
H^{\pm}_2=\frac{w\widetilde{\phi}_1^{\pm}+v_1 \phi^{\pm}_3  }{\sqrt{v_1^2 + w^2}},\qquad m^2_{H^{\pm}_2}=\frac{(v_1^2+w^2)(v_2\mu_{\phi}+v_1w\tilde{\lambda}_{13})}{2v_1w}.
\end{equation}

Finally, the only charged scalar with $B-L= 1$, $\phi^{+}_4$, remains unmixed after spontaneous symmetry breaking, and gets the squared mass
\begin{equation}
m^2_{\phi_4^{\pm}}=\frac{1}{2}\left[
v_1^2(\lambda_{14}+\tilde{\lambda}_{14})+v_2^2\lambda_{24}+w^2 \lambda_{34}+2 \mu_4^2 \right].
\end{equation}

\section{Gauge Sector}
\label{sec:gauge}

In this section, we study the vector boson spectrum of the extended electroweak sector which contains ten gauge fields.
After spontaneous symmetry breaking, gauge boson masses are generated, as usual, through the terms
$\mathcal{L}\supset (D_\mu \Phi_i)^\dagger(D^\mu \Phi^i)$, where the covariant derivative acting on the scalar anti-triplets is defined as
\begin{equation}
	D_\mu \Phi_i= \left[ \partial_\mu + i g_L \frac{\lambda^{a\,*}}{2}W^{a}_{\mu}- i g_X X B_{\mu}-i g_N N C_{\mu}\right]\Phi_i = \left(\partial_\mu+ i \frac{g_L}{2}\mathcal{P}_\mu\right)\Phi_i
\end{equation}
where $W^{a}_\mu$ are the gauge fields of $SU(3)_L$, $\lambda^{a}$ are the Gell-Mann matrices, $B_\mu$ is the gauge field of $U(1)_X$, and $C_{\mu}$ is the gauge field of  $U(1)_N$ and 
\begin{eqnarray}
\mathcal{P}_\mu=
	\begin{pmatrix}
	W^3+\frac{W^8}{\sqrt{3}} - 2\left(t_X X B + t_N N C \right) & \sqrt{2}W^- & \sqrt{2}W^{\prime-} \\ \sqrt{2} W^+ & -W^3+\frac{W^8}{\sqrt{3}} - 2\left(t_X X B + t_N N C\right) & \sqrt{2}X^{0*} \\
	\sqrt{2} W^{\prime+} & \sqrt{2} X^{0} & -2\left(\frac{W^8}{\sqrt{3}} + t_X X B + t_N N C\right) \end{pmatrix}_\mu,
\end{eqnarray}
with
\begin{equation}\label{cgb}
 W_\mu^\pm = \frac{W_\mu^1\mp i W_\mu^2}{\sqrt{2}}\,,\,\,\, W_\mu^{\prime \pm}=\frac{W_\mu^4\mp i W_\mu^5}{\sqrt{2}}\,,\,\,\, X_\mu^{0(*)} =\frac{W_\mu^6\mp i W_\mu^7}{\sqrt{2}}\,.
\end{equation}
In addition, we assume another source for gauge boson masses through the Stueckelberg mechanism for the Abelian $U(1)_N$ symmetry \cite{Ruegg:2003ps}.
The masses and states of the ten electroweak gauge bosons are discussed below.

\subsection{Neutral gauge bosons and Stueckelberg mechanism}

After spontaneous symmetry breaking, the two gauge bosons of the abelian symmetries, $B_\mu$ and $C_\mu$, and the two fields associated with the diagonal generators of $SU(3)_L$, $W_\mu^3$ and $W_\mu^8$, mix among themselves. Assuming the kinetic mixing between the gauge bosons $B_\mu$ and $C_\mu$ can be neglected~\footnote{The effects of non-vanishing kinetic mixings have been discussed in Refs.~\cite{Holdom:1985ag,Feldman:2007wj,Williams:2011qb}.}, the relevant terms contributing to the neutral boson masses, written in the basis $\mathcal{B}^{T}_{\mu}=(W^{3}_{\mu},W^{8}_{\mu},B_{\mu},C_{\mu})$, are
\begin{equation}\label{St1}
\mathcal{L}\supset\frac{1}{2}\mathcal{B}^{T}_{\mu}M^2_0 \mathcal{B}^{\mu}+\frac{1}{2}(m C^\mu-\partial^\mu \sigma)^2+\mathcal{L}_{\mathrm{gf}}^{\mathrm{St}}.
\end{equation} 
Here, $M^2_0$ is the squared mass matrix coming from the Higgs mechanism, $m$ is the Stueckelberg mass of the $C^\mu$ gauge field, and $\sigma$ is the scalar Stueckelberg compensator that renders the second term in Eq. (\ref{St1}) invariant under the gauge transformations,
\begin{equation}
\begin{split}
C^\mu&\rightarrow C^\mu+\partial^\mu\Omega(x),\\
\sigma&\rightarrow \sigma+m \Omega(x),
\end{split}
\end{equation}
with an arbitrary spacetime function $\Omega(x)$. The gauge fixing term $\mathcal{L}_{\mathrm{gf}}^{\mathrm{St}}$ can be chosen as
\begin{equation}
\mathcal{L}_{\mathrm{gf}}^{\mathrm{St}}=-\frac{1}{2\xi} \left\{\partial^\mu C_\mu+\xi\left[m\sigma-\frac{2}{3} g_N \left(\sqrt{v_1^2+v_2^2} G_1+2\sqrt{v_1^2+w^2} G_2\right)\right]\right\}^2,
\end{equation}
ensuring (up to a total derivative) that the gauge field $C^\mu$ decouples from the gradients $\partial^\mu \sigma$, $\partial^\mu G_1$  and $\partial^\mu G_2$.
Notice that after gauge fixing, the Lagrangian is still invariant under a restricted set of gauge functions $\Omega(x)$, subject to the same equation of motion as $\sigma$, {\it i.e.}
$(\partial^2+\xi m^2)\Omega=(\partial^2+\xi m^2)\sigma=0$. This dynamical restriction guarantees the propagation of three degrees of freedom for the massive vector field $C_\mu$.
Moreover, $\mathcal{L}_{\mathrm{gf}}^{\mathrm{St}}$ introduces a mixing between the scalars $\sigma$, $G_1$ and $G_2$.

After implementing the Stueckelberg mechanism outlined above, the squared-mass matrix of the neutral gauge bosons becomes
\begin{equation}
M^2=\frac{g_L^2}{2}\left(
\begin{array}{cccc}
 \frac{1}{2} \left(v_1^2+v_2^2\right)  & \frac{v_1^2-v_2^2 }{2 \sqrt{3}} & -\frac{1}{3} \left(2 v_1^2+v_2^2\right)  t_X & \frac{2}{3} \left(v_2^2-v_1^2\right) t_N \\
 \frac{v_1^2-v_2^2 }{2 \sqrt{3}} & \frac{1}{6} \left(v_1^2+v_2^2+4 w^2\right) &  \frac{\left(v_2^2-2 v_1^2-2 w^2\right)  t_X}{3 \sqrt{3}}& -\frac{2 \left(v_1^2+v_2^2+4 w^2\right) t_N}{3 \sqrt{3}} \\
 -\frac{1}{3} \left(2 v_1^2+v_2^2\right)  t_X & \frac{\left(v_2^2-2 v_1^2-2 w^2\right)  t_X}{3 \sqrt{3}} & \frac{2}{9} \left(4 v_1^2+v_2^2+w^2\right)  t_X^2 & \frac{4}{9} \left(2 v_1^2-v_2^2+2 w^2\right)  t_N t_X \\
 \frac{2}{3} \left(v_2^2-v_1^2\right)  t_N & -\frac{2 \left(v_1^2+v_2^2+4 w^2\right)  t_N}{3 \sqrt{3}} & \frac{4}{9} \left(2 v_1^2-v_2^2+2 w^2\right)  t_N t_X & \frac{2}{g_L^2}m^2+\frac{8}{9} \left(v_1^2+v_2^2+4 w^2\right) t_N^2 \\
\end{array}
\right),
\end{equation}
with $t_X=g_X/g_L$ and $t_N=g_N/g_L$.
In order to diagonalize $M^2$, several changes of basis will be required. In this analysis we follow the procedure described in \cite{Dong:2014wsa}.

We first identify the photon field $A_\mu$.
The transformation matrix to the basis $(A_\mu,Z_{1\mu},Z^{\prime}_{1\mu},C_\mu)$ is given by
\begin{equation}
\left(\begin{array}{c}
A_\mu\\Z_{1\mu}\\Z^{\prime}_{1\mu}\\C_\mu
\end{array}\right)=U_1\left(\begin{array}{c}
W^{3}_{\mu}\\W^{8}_{\mu}\\B_{\mu}\\C_{\mu}
\end{array}\right),
\qquad U_1=\left(
\begin{array}{cccc}
 \frac{\sqrt{3} t_X}{\sqrt{4 t_X^2+3}} & \frac{t_X}{\sqrt{4 t_X^2+3}} & \frac{\sqrt{3}}{\sqrt{4 t_X^2+3}} & 0 \\
 \sqrt{\frac{t_X^2+3}{4 t_X^2+3}} & -\frac{\sqrt{3} t_X^2}{\sqrt{\left(t_X^2+3\right) \left(4 t_X^2+3\right)}} & -\frac{3 t_X}{\sqrt{\left(t_X^2+3\right) \left(4 t_X^2+3\right)}} & 0 \\
 0 & \frac{\sqrt{3}}{\sqrt{t_X^2+3}} & -\frac{t_X}{\sqrt{t_X^2+3}} & 0 \\
 0 & 0 & 0 & 1 \\
\end{array}
\right),
\end{equation}
such that
\begin{eqnarray}
M^{\prime 2}&=&U_1 M^2 U_1^T= \left(
\begin{array}{cc}
0 & 0 \\
0 & M^{\prime 2}_s
\end{array}
\right),
\end{eqnarray}
with
\begin{eqnarray}
&M^{\prime 2}_s&=\frac{g_L^2}{2}\times\\
&  &\left(
\begin{array}{ccc}
 \frac{\left(v_1^2+v_2^2\right)  \left(4 t_X^2+3\right)}{2 \left(t_X^2+3\right)} & \frac{\sqrt{4 t_X^2+3} [v_1^2(4t_X^2+3)+v_2^2(2t_x^2-3)]}{6 \left(t_X^2+3\right)} & -\frac{2}{3}(v_1^2-v_2^2) t_N \left(\frac{4t_X^2+3}{t_X^2+3}\right)^{1/2} \\
 \frac{\sqrt{4 t_X^2+3} [v_1^2(4t_X^2+3)+v_2^2(2t_X^2-3)]}{6 \left(t_X^2+3\right)} & \frac{v_1^2(3+4t_X^2)^2+v_2^2(3-t_X^2)^2+4w^2(3+t_X^2)^2}{18 \left(t_X^2+3\right)} & -\frac{2 t_N [2\left(2 v_1^2- v_2^2+2 w^2\right) t_X^2+3 \left(v_1^2+v_2^2+4 w^2\right)]}{9 \sqrt{t_X^2+3}} \\
 -\frac{2}{3} (v_1^2-v_2^2)  t_N \left(\frac{4t_X^2+3}{t_X^2+3}\right)^{1/2} & -\frac{ 2 t_N [2\left(2 v_1^2- v_2^2+2 w^2\right) t_X^2+3 \left(v_1^2+v_2^2+4 w^2\right)]}{9 \sqrt{t_X^2+3}} & \frac{2 m^2}{g_L^2 }+\frac{8}{9} \left(v_1^2+v_2^2+4 w^2\right) t_N^2 \\
\end{array}
\right).\nonumber
\end{eqnarray}
Therefore, the photon is identified as
\begin{equation}
A_\mu= \frac{1}{\sqrt{4 t_X^2+3}}\left(\sqrt{3} t_X W^{3}_{\mu}+ t_X W^{8}_{\mu}+\sqrt{3} B_{\mu}\right).
\end{equation}
For the second diagonalization to the basis $(A_\mu,Z_\mu, Z^{\prime}_{2\mu},C^{\prime}_\mu)$, we use a ``seesaw approximation''~\cite{Schechter:1981cv} \\
\begin{equation}
U_2 \approx 
\left(
\begin{array}{cccc}
 1 & 0 & 0 & 0 \\
 0 & 1 & \varepsilon _1 & \varepsilon _2 \\
 0 & -\varepsilon _1 & 1 & 0 \\
 0 & -\varepsilon _2 & 0 & 1 \\
\end{array}
\right),
\end{equation}
where $\varepsilon_1$ and  $\varepsilon_1$ are the two components of a small vector given by
\begin{eqnarray}
\varepsilon &\equiv&- (m^2_{Z_1Z_1^{\prime}}, \, m^2_{Z_1C})
\left(
\begin{array}{cc}
m^2_{Z^\prime_1} & m^2_{Z^{\prime}_1C} \\
 m^2_{Z^{\prime_1}C} & m^2_C\\
\end{array}
\right)^{-1},\\
\varepsilon_1&=&-\frac{\sqrt{4 t_X^2+3} \left\lbrace 2 t_X^2 \left[8 g_L^2 t_N^2 \left(w^2v_1^2+v_2^2w^2+v_1^2v_2^2\right)+3 m^2 \left(2 v_1^2+v_2^2\right)\right]+9 m^2 (v_1^2-v_2^2)\right\rbrace}{ 4 t_X^4 \left[ 4 g_L^2 t_N^2 \left(w^2v_1^2+v_2^2w^2+v_1^2v_2^2\right)+m^2 \left(4 v_1^2+v_2^2+w^2\right)\right] +3 m^2\left[4 t_X^2 \left(2 v_1^2-v_2^2+2 w^2\right)+3 \left(v_1^2+v_2^2+4 w^2\right)\right]},\nonumber\\
\varepsilon_2&=&- \frac{4 g_L^2 t_N t_X^2 \sqrt{t_X^2+3} \sqrt{4 t_X^2+3} \left(v_1^2 \left(v_2^2+w^2\right)+v_2^2 w^2\right)}{ 4 t_X^4 \left\lbrace 4 g_L^2 t_N^2 \left[w^2 \left(v_1^2+v_2^2\right)+v_1^2 v_2^2\right]+m^2 \left(4 v_1^2+v_2^2+w^2\right)\right\rbrace +3 m^2\left[4 t_X^2 \left(2 v_1^2-v_2^2+2 w^2\right)+3 \left(v_1^2+v_2^2+4 w^2\right)\right]},\nonumber
\end{eqnarray}
which are suppressed by the hierarchy $v_1,v_2<<w,m$.

Then, after the second diagonalization, we have
\begin{equation}
M^{\prime \prime 2}=U_2 M^{\prime 2} U_2^T= \left(
\begin{array}{ccc}
 0 & 0 & 0 \\
 0 & m_{Z}^2 & 0 \\
 0 & 0 & M^{\prime \prime 2}_s \\
\end{array}
\right),
\end{equation}
with
\begin{equation}
\begin{split}
m_{Z}^2 &\approx m_{Z_1}^2 +2 \left(\varepsilon_1 , \varepsilon_2
\right)
 \left( \begin{array}{c}
m^2_{Z_1Z_1^\prime}\\
m^2_{Z_1C}\\
\end{array}
\right)\\
&\approx  \frac{ g_L^2\left(v_1^2+v_2^2\right) \left(4 t_X^2+3\right)}{4 \left(t_X^2+3\right)},
\end{split}
\end{equation}
which can be identified with the squared mass of the physical electroweak $Z_\mu$ boson and 
\begin{equation}
M_s^{\prime \prime 2} \approx  
 \left( \begin{array}{cc}
m^2_{Z_2^\prime} & m^2_{Z_2^{\prime} C^{\prime}}\\
m^2_{Z_2^{\prime}C^{\prime}} & m^2_{C^{\prime}} \\
\end{array}
\right).
\end{equation}
Finally we can diagonalize $M_s^{\prime \prime 2}$  to the $(A_\mu,Z_\mu,Z^{\prime}_\mu,Z^{\prime\prime}_\mu)$ basis through
\begin{equation}
U_3=\left(
\begin{array}{cccc}
1 & 0 & 0 & 0 \\
0 & 1 & 0 & 0 \\
0 & 0 & c_{\zeta} & s_{\zeta}\\
0 & 0 & -s_{\zeta} & c_{\zeta}\\
\end{array}
\right),
\end{equation}
and the diagonal squared mass matrix for the physical gauge bosons becomes 
\begin{equation}
M^{\prime \prime \prime 2}= U_3 M^{\prime \prime 2} U_3^T\,,
\end{equation}
where the mixing angle is given by
\begin{equation}
\tan2\zeta\approx\frac{8 w^2 g_L^2 t_N \sqrt{t_X^2+3}}{w^2 g_L^2 \left(16 t_N^2-t_X^2-3\right)+9 m^2},
\end{equation}
and the diagonal entries can be identified as the squared masses for the physical $Z^\prime$ and $Z^{\prime\prime}$ bosons,
\begin{eqnarray}\label{Zpsmass}
 m_{Z^\prime,\, Z^{\prime\prime}}^2 &=& \frac{1}{18} \left\{w^2 g_L^2 \left(16
   t_N^2+t_X^2+3\right)+9 m^2\mp\mathcal{G}\sqrt{\left[w^2 g_L^2 \left(16 t_N^2-t_X^2-3\right)+9 m^2\right]{}^2+64 w^4 g_L^4 t_N^2 \left(t_X^2+3\right)}\right\}\,,\end{eqnarray}
with $\mathcal{G}=\mathrm{sign}[w^2 g_L^2 \left(16 t_N^2-t_X^2-3\right)+9 m^2]$. 

\subsection{Complex neutral gauge bosons}

The complex gauge boson $X^0_\mu$, with $B-L=2$, does not mix with the other neutral vector fields.
After spontaneous symmetry breaking, $X^0_\mu$, whose associated would-be Goldstone boson is $G_3$ in Eq. (\ref{G3}), gets the following mass term
\begin{equation}
m^2_{X^0}=\frac{g_L^2}{4} \left(v_2^2+w^2\right).
\end{equation}

\subsection{Charged gauge bosons}

The charged gauge bosons present in the model, $W_\mu^\pm$ and $W_\mu^{\prime\pm}$, become massive after electroweak symmetry breaking but do not mix due to their different $B-L$ charges.

The first mass eigenstate is identified with the charged \sm electroweak W boson, whose would-be Goldstone bosons given by $G_4^{\pm}$, and has the squared mass,
\begin{equation}
m^2_W=\frac{g_L^2}{4}  \left(v_1^2+v_2^2\right).
\end{equation}
Finally, the other charged gauge boson is heavy and eats up the complex would-be Goldstone boson $G_5^{\pm}$ in order to acquire the squared mass
\begin{equation}
m^2_{W^{\prime}}=\frac{g_L^2}{4}  \left(v_1^2+w^2\right).
\end{equation}
To sum up we note that, despite the conservation of $B-L$, all of the gauge bosons acquire adequate masses through the interplay of the standard Higgs mechanism with the Stueckelberg mechanism, leaving only the photon massless, as in the Standard Model.
In particular, we would like to reinforce the importance of the Stueckelberg mechanism which provides the $B-L$ gauge boson with a mass, while keeping the associated symmetry fully preserved. 
As mentioned above, the conservation of the $B-L$ symmetry, not affected by the Stueckelberg mechanism, is what ensures two appealing features of our model, namely, the Dirac nature of neutrinos and the stability of dark matter.

\section{Charged fermions}
\label{sec:fermion}

The Yukawa interactions invariant under all the defining symmetries of the model are
\begin{eqnarray}\label{lagY3}
 -\mathcal{L}_{\rm Yuk} &=& \,
 y^{e}_{ab}\,\overline{e^{a}_{R} }  \,\Phi_1^\dagger \psi^{b}_L +y^{S}_{ab} \,\overline{S^{a}_{R}}\,\Phi_4^{\dagger} \,  \psi^{b}_L +
 \tfrac{M^{ S}_{ab}}{2} \overline{(S^{a}_R)^c}  S^{b}_R  \nonumber\\
 &+&  
y^{u}_{a \alpha}\, \overline{u^{a}_R}\,\Phi_1^{T} \, Q_L^{\alpha}  +  
y^{u}_{a 3}\, \overline{u^{a}_R}\,\Phi_2^{\dagger} \,Q_L^{3}  
+y^{d}_{a 3} \,\overline{d^{a}_R}\,\Phi_1^{\dagger} \, Q_L^{3} + 
y^{d}_{a\alpha } \, \overline{d^{a}_R}\,\Phi_2^{T} \,Q_L^{\alpha}  \nonumber \\
&+& 
 y^{U}_{33}\, \overline{U^3_R} \,\Phi_3^{\dagger}  \,Q_L^{3}
  +  y^{D}_{\alpha\beta} \, \overline{D_R^{\alpha}}\,\Phi_3^{T} Q_L^{\beta}\,  
 + \mathrm{h.c.}\,.
\end{eqnarray}
After spontaneous symmetry breaking, the above interactions lead to the following mass matrices for the fermions:
\begin{itemize}
\item Charged leptons:
\begin{equation}
M_{ab}^{e}=y_{ab}^{e}\frac{v_1}{\sqrt{2}}.
\end{equation}
\item Up-type quarks, basis $(u,c,t,U_3)$:
\begin{equation}\label{uqmass}
M_{u}=\frac{1}{\sqrt{2}}\left(
\begin{array}{cccc}
 v_1 y^u_{11} & v_1  y^u_{12} &  -v_2  y^u_{13} & 0 \\
 v_1 y^u_{21} &  v_1 y^u_{22} &  -v_2  y^u_{23} & 0 \\
 v_1 y^u_{31} &  v_1  y^u_{32} &  -v_2  y^u_{33} & 0 \\
0 & 0 & 0 & w y^{U}_{33} \\
\end{array}
\right).
\end{equation}
\item Down-type quarks, basis $(d,s,b,D_{1},D_{2})$:
\begin{equation}\label{dqmass}
\begin{split}
M_{d}&=\frac{1}{\sqrt{2}}\left(
\begin{array}{ccccc}
 v_2 y^d_{11} & v_2  y^d_{12} &  v_1  y^d_{13} & 0 &0\\
  v_2 y^d_{21} &   v_2 y^d_{22} &  v_1  y^d_{23} & 0 &0\\
  v_2  y^d_{31} &   v_2  y^d_{32} &  v_1  y^d_{33} & 0 &0\\
0 & 0 & 0 &  w y_{11}^{D} & w y_{12}^{D} \\
0 & 0 & 0 &  w y_{12}^{D} &  w y_{22}^{D} \\
\end{array}
\right)\,.
\end{split}
\end{equation}
\end{itemize}
Realistic quark masses can be easily obtained from the above mass matrices, as the \sm and exotic quarks remain unmixed by virtue of the unusual $B-L$ charges of the exotic sector.
This is reflected by the block-diagonal form of the above matrices, which also implies the unitarity of the Cabibbo-Kobayashi-Maskawa matrix describing quark mixing.

Notice that, from the above Yukawa interactions, neutrinos remain massless at the tree level. 

\section{Scotogenic Neutrino Masses}
\label{sec:neutrino}

\begin{figure}[htbp]
\begin{center}
\includegraphics[scale=1.2]{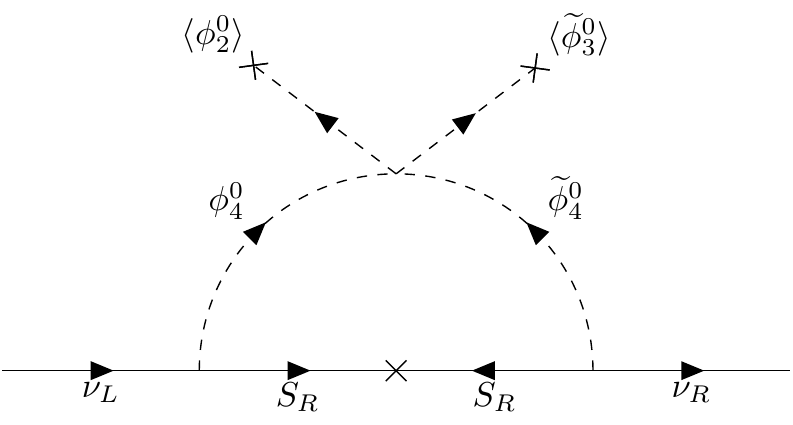}
\caption{One-loop scotogenic Dirac neutrino mass generation mechanism.}
\label{Neutrino_diagram}\label{fig}
\end{center}
\end{figure}

As previously shown, in the present model the gauged $B-L$ symmetry remains unbroken and so does the matter parity $M_P$.
Furthermore, the $\mathrm{U(1)_{PQ}}$ symmetry, only broken softly in the scalar sector, forbids the appearance of a tree-level neutrino-mass-giving Yukawa term.
However, the Yukawa interactions in Eq. (\ref{lagY3}) allow the emergence of a calculable one-loop contribution to the neutrino masses via the diagram in Fig. \ref{Neutrino_diagram}.

Assuming that the Majorana mass of the fermion singlets $S_R$ is already diagonal $M^S=\mathrm{diag}(M_1,M_2,M_3)$, the neutrino mass matrix generated by the scotogenic loop in the basis $(\nu_L,(\nu_{R})^{c})$ reads
\begin{equation}
M_{\nu}=\left(
\begin{array}{cc}
0 &m_{\nu}  \\
m_{\nu}^{T} &0  \\
\end{array}
\right),
\end{equation}
with neutrino Dirac masses
\begin{equation}
(m_{\nu})_{ab}=\sum_{k=1}^{3}\frac{M_k y^{S}_{ka}y^{S}_{kb} \sin2\theta}{32\pi^2}\left[\frac{m^2_{\eta_1}}{m^2_{\eta_1}-M_k^2}\ln\frac{m^2_{\eta_1}}{M_k^2}-\frac{m^2_{\eta_2}}{m^2_{\eta_2}-M_k^2}\ln\frac{m^2_{\eta_2}}{M_k^2}\right].
\end{equation}
Notice that from Eq. (\ref{tmix}), if the relevant quartic couplings are of the same order, the angle $\theta$ is already suppressed, of $\mathcal{O}(v_2/w)$.
Besides, the internal fields in the loop are odd under $M_P$, while the \sm fields are even.
Thus, the lightest $M_P$-odd field is automatically stable and, if it is electrically neutral, can be identified as a dark matter candidate.
In our model, the stable dark matter candidate will be the lightest field among the complex scalars $\eta_i$ and Majorana fermions $S_{aR}$.

\section{Dark Matter}
\label{sec:DM}

In order to illustrate the viability of our model as a theory of dark matter, we study a simplified scenario in which all the non-SM fields are heavy and decouple,
except for the complex scalars $\eta_1$ and $\eta_2$. In this case, only the Higgs and the $Z$-boson portals are available.
In general, the region of the parameters compatible with the observed relic abundance and direct dark matter detection experiments is very constrained for a complex scalar, unless co-annihilation takes place due to $\eta_1$ and $\eta_2$ being almost degenerate \cite{Kakizaki:2016dza}.
Besides, consistency with direct detection experiments requires the coupling between the complex dark
matter candidate and the $Z$-boson to be very small.
This can be easily achieved in our model since the mixing angle $\theta$ in Eq. (\ref{tmix}) is naturally
of $\mathcal{O}(v_2/w^{-1})$. We assume that $\eta_1$ is our dark matter candidate,
composed mostly by the $\mathrm{SU(2)_L}$ singlet $\widetilde{\phi}^0_4$, and couples to the $Z$-boson only through its suppressed mixing with $\eta_2$. 

\begin{figure}[H]
\centering
\includegraphics[width=0.7\textwidth]{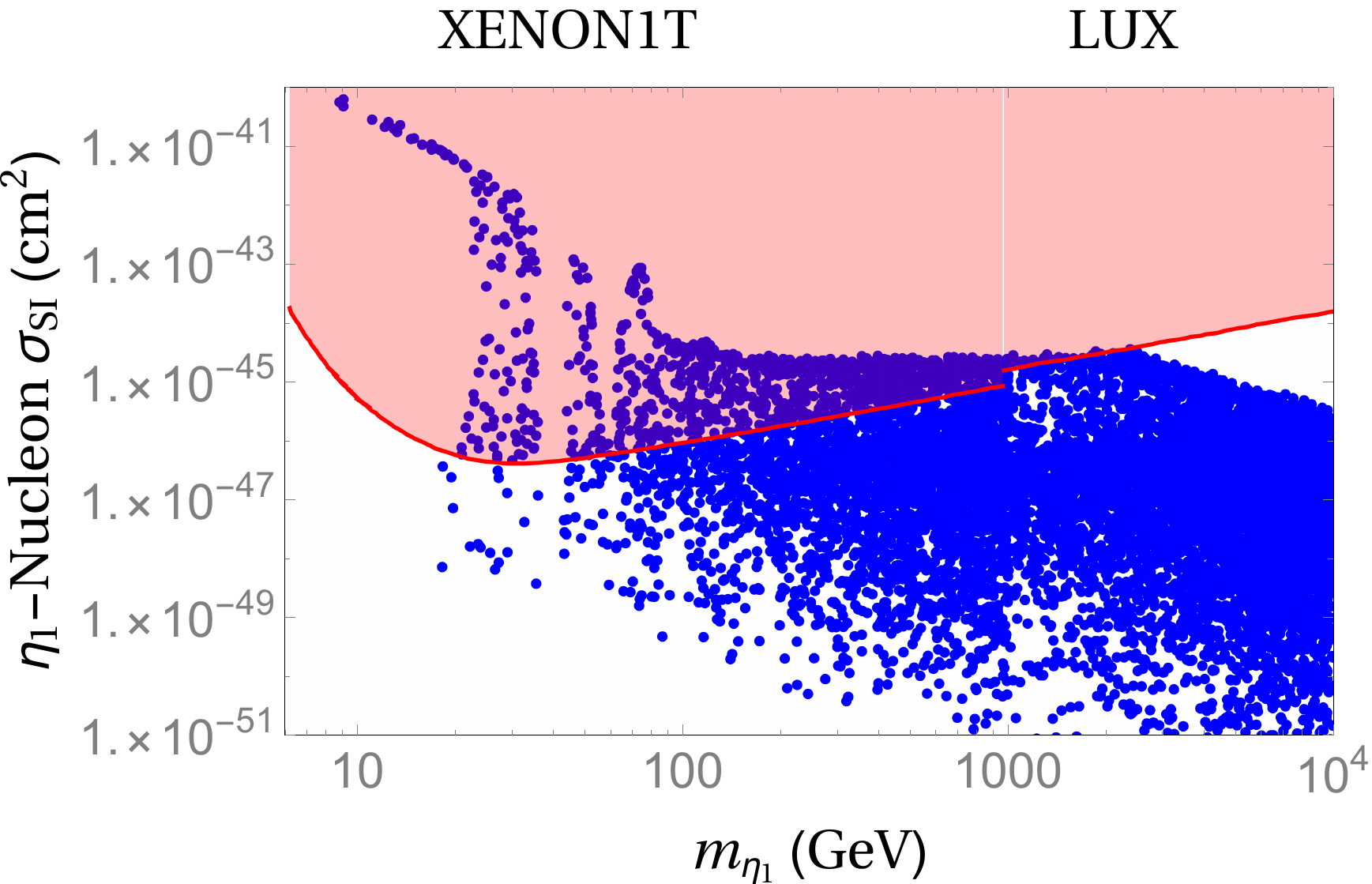}
\caption{The direct detection and relic abundance constraints on the dark matter mass $m_{\eta_1}$. Each blue point corresponds to a solution $(\lambda,\theta,m_{\eta_1},m_{\eta_2})$ with the correct relic abundance $\Omega h^2=0.120$. See the text for details. The red shaded region is ruled out by direct detection experiments, XENON1T \cite{Aprile:2018dbl} and LUX \cite{Akerib:2016vxi}. Notice that the stronger constraint below $1\,\mathrm{TeV}$ comes from XENON1T and above $1\,\mathrm{TeV}$ from LUX.}
\label{DMfig}
\end{figure}

From our previous analysis of the scalar spectrum, the condition $m_{\eta_1}<m_{\eta_2}$ in Eq. (\ref{meta}) translates to $\tilde{\lambda}_{24} v_2^2-\tilde{\lambda}_{34} w^2>0$, easily achieved by a natural negative value of $\tilde{\lambda}_{34}$. Defining
\begin{equation}
\begin{split}
\mu_S^2&\equiv \frac{1}{2}\left[w^2(\lambda_{34}+\tilde{\lambda}_{34})+2\mu^2_4\right],\\
\mu_D^2&\equiv \frac{1}{2}\left[v_2^2\tilde{\lambda}_{24}+w^2 \lambda_{34}+2\mu^2_4\right],
\end{split}
\end{equation}
in Eq. (\ref{MsqS}), one can eliminate these scales and $v_2 w \lambda'$ in terms of the physical masses $m_{\eta_1}$, $m_{\eta_2}$ and the mixing angle $\theta$ as
\begin{equation}
\begin{split}
&v_2 w \lambda'=2 \left(m_{\eta_2}^2-m_{\eta_1}^2\right) \sin 2 \theta, \\
\mu_S^2&=m_{\eta_2}^2 \sin ^2\theta +m_{\eta_1}^2 \cos ^2\theta -\frac{1}{2}
   \lambda _{14} v_1^2-\frac{1}{2} \lambda _{24} v_2^2,\\
\mu_D^2&=m_{\eta_1}^2 \sin
   ^2\theta +m_{\eta_2}^2 \cos ^2\theta -\frac{1}{2} \lambda _{14} v_1^2-\frac{1}{2} \lambda
   _{24} v_2^2.
\end{split}
\end{equation}
We have studied the relic abundance and direct detection constraints for this scenario, setting for simplicity $v_1=v_2=v_{EW}/\sqrt{2}$ and $ \lambda _{14}= \lambda _{24}=\lambda$ with vanishing non-relevant couplings. In our analysis we have varied randomly the relevant parameters in the ranges $0<|\lambda|<1$, $0<|\theta|<0.01$, $0<m_{\eta_1}<10^4\, \text{GeV}$ and $m_{\eta_1}<m_{\eta_2}<1.1m_{\eta_1}$. The results are shown in Fig. \ref{DMfig}, where each blue point corresponds to a solution $(\lambda,\theta,m_{\eta_1},m_{\eta_2})$ in parameter space complying with the correct relic abundance $\Omega h^2=0.120$ \cite{Aghanim:2018eyx}. One can see that the model contains plenty of parameter combinations well below the current direct detection bounds,
but also within the sensitivity of the current experiments, like Xenon1T.
Before ending this section we wish to remark that Fig. \ref{DMfig} is plotted for a simplified scenario in which only the Higgs and $Z$-boson portals are available.
This need not be the case. In our model, the allowed parameter space can be considerably richer due to the presence of Majorana fermions, like $S_{aR}$, providing new channels
for dark matter annihilation. Similarly, more parameter combinations become available when the vector bosons $Z^\prime$ and $Z^{\prime\prime}$, whose masses are given in Eq. (\ref{Zpsmass}),
  are active in mediating dark matter annihilation processes, instead of simply decoupled, as assumed in the above example. A dedicated study lies outside the scope of this paper.

\section{Summary and conclusions}
\label{sec:Conclusions}

In this work we have proposed a simple scotogenic extension of the original Singer-Valle-Schechter 3-3-1 model in which neutrinos are Dirac fermions as a result of a conserved $B-L$ gauge symmetry.
In such minimal SVS gauge extension of the \sm neutrino masses arise through the radiative exchange of the simplest scotogenic ``dark''  sector, as indicated by the diagram in Fig. \ref{fig}. 
Conservation of $B-L$ gauge symmetry in the \TrTrOO theory ensures the stability of dark matter, linked to the Dirac nature of neutrinos.
By combining the Higgs and the Stueckelberg mechanisms, one ensures that all neutral gauge bosons acquire adequate non-zero masses.
Our present construction bears similarities with that in Ref.~\cite{Leite:2020wjl}, but within a richer theoretical framework.
Indeed, the present one also touches other \sm shortcomings, such as the existence of three families, which emerges just from anomaly cancellation.
Stable dark matter is ``predestined''~\cite{Ma:2018bjw}, in the sense that the imposition of additional symmetries is not required.
We have given a detailed study of the basic structure of the theory.
For example, we noted that due to our quantum numbers we have block-diagonal quark mass matrices, Eqs.~(\ref{uqmass}) and (\ref{dqmass}), implying the unitarity of the CKM matrix describing quark mixing.
This implies that the new neutral gauge bosons can have flavor-changing interactions at the tree level, as in the SVS model.
These arise from the underlying structure of the neutral current dictated by the anomaly cancellation.
As a result, in addition to direct searches through dilepton studies at the LHC, heavy neutral gauge bosons induce mass differences in neutral meson systems.
These can lead to observable phenomena if they lie within the few TeV scale.
For example, for $v_1 \sim v_2 $ GeV  if one takes $m\to\infty$, $w \sim 10^4$ GeV as a benchmark, one finds that the $B-L$ Stueckelberg gauge boson decouples, leaving adequate masses for the other new intermediate gauge bosons, around $4$ TeV, consistent with current limits from flavor changing neutral current and searches at the LHC run 2 at $13$ TeV~\cite{Queiroz:2016gif}.
  Likewise, one can check that the scalar masses expected, e.g. from Eqs. (\ref{Hmass}) and (\ref{pseudo2}), are also phenomenologically viable.
  The same happens for finite values of the Stueckelberg gauge boson mass parameter: in this case, one also obtains gauge boson mass values in agreement with current limits.
 We expect, however, that they can lie within the sensitivities expected, for example, at the High Luminosity-LHC, LHCb as well as upcoming B factories.

 Concerning the dark matter content of our model, in section \ref{sec:DM} we have analised the case for a complex dark matter scalar candidate. For definiteness we took a simple scenario where only the Higgs and $Z$-boson portals are available.
We have shown that, even in this simplified scenario, there are parameter combinations that accomodate the correct dark matter relic abundance in agreement with direct detection constraints.
The viable parameter space is expected to be substantially widened when other channels for dark matter annihilation are taken into account, {\it e.g.} those mediated by the vector bosons $Z^{\prime}$ and $Z^{\prime\prime}$.  \\[.4cm]

Last, but not least, we stress that, in contrast to previous 3-3-1-1 models, here neutrinos get radiative scotogenic Dirac masses, rather than Majorana masses from the conventional seesaw mechanism.
A discovery of neutrinoless double beta decay would therefore invalidate our present construction.

\acknowledgements 
\noindent

Work supported by the Spanish grants FPA2017-85216-P (AEI/FEDER, UE), PROMETEO/2018/165 (Generalitat Valenciana) and the Spanish Red Consolider MultiDark FPA2017-90566-REDC. J. L. acknowledges financial support under grant 2019/04195-7, S\~ao Paulo Research Foundation (FAPESP).  A.M. acknowledges support by CONACyT. CAV-A is supported by the Mexican C\'atedras CONACYT project 749 and SNI 58928. Numerical work performed in GuaCAL (Guanajuato Computational Astroparticle Lab). The relic abundance and
direct detection constraints are calculated using the MicroOmegas package \cite{Belanger:2018ccd}.

\bibliographystyle{utphys}
\bibliography{bibliography}
\end{document}